\documentclass[sigconf]{acmart} %authordraft,anonymous,
\usepackage{multirow}
\makeatletter
\newcommand{\thickhline}{%
    \noalign {\ifnum 0=`}\fi \hrule height 0.7pt
    \futurelet \reserved@a \@xhline
}
\DeclareRobustCommand*\cal{\@fontswitch\relax\mathcal}

\AtBeginDocument{%
  \providecommand\BibTeX{{%
    \normalfont B\kern-0.5em{\scshape i\kern-0.25em b}\kern-0.8em\TeX}}}

\setcopyright{acmcopyright}
\copyrightyear{2018}
\acmYear{2018}
\acmDOI{10.1145/1122445.1122456}

\acmConference[Woodstock '18]{Woodstock '18: ACM Symposium on Neural
  Gaze Detection}{June 03--05, 2018}{Woodstock, NY}
\acmBooktitle{Woodstock '18: ACM Symposium on Neural Gaze Detection,
  June 03--05, 2018, Woodstock, NY}
\acmPrice{15.00}
\acmISBN{978-1-4503-XXXX-X/18/06}

\begin{document}

%%
%% The "title" command has an optional parameter,
%% allowing the author to define a "short title" to be used in page headers.
\title{A Non-sequential Approach to Deep User Interest Model for CTR Prediction}

%%
%% The "author" command and its associated commands are used to define
%% the authors and their affiliations.
%% Of note is the shared affiliation of the first two authors, and the
%% "authornote" and "authornotemark" commands
%% used to denote shared contribution to the research.

\author{Keke Zhao}
\authornote{Corresponding authors.}
\affiliation{%
  \institution{Antgroup}
  \streetaddress{ 556  Xixi Rd}
  \city{Hang Zhou}
  \state{Zhejiang}
  \country{China}}
\email{keke.zhaokk@antgroup.com}

\author{Xing Zhao}
\authornotemark[1]
\affiliation{%
  \institution{Antgroup}
  \streetaddress{ 556  Xixi Rd}
  \city{Hang Zhou}
  \state{Zhejiang}
  \country{China}}
\email{xing1zhao4@yahoo.com}

\author{Qi Cao}
\affiliation{%
  \institution{Antgroup}
  \streetaddress{ 556  Xixi Rd}
  \city{Hang Zhou}
  \state{Zhejiang}
  \country{China}}
\email{qili.cq@antgroup.com}

\author{Linjian Mo}
\affiliation{%
  \institution{Antgroup}
  \streetaddress{ 556  Xixi Rd}
  \city{Hang Zhou}
  \state{Zhejiang}
  \country{China}}
\email{linyi01@antgroup.com}

%%
%% By default, the full list of authors will be used in the page
%% headers. Often, this list is too long, and will overlap
%% other information printed in the page headers. This command allows
%% the author to define a more concise list
%% of authors' names for this purpose.

% \renewcommand{\shortauthors}{Keke Zhao and Xing Zhao, et al.}

%%
%% The abstract is a short summary of the work to be presented in the
%% article.

\begin{abstract}
% 曹琦
%Click-Through Rate (CTR) prediction plays an important role in many industrial applications. The existing methods dedicate to improve CTR performance by mining interest from user behavior sequences. However, there are two important problems with these approaches: 1). It employs fixed length to represent behavior sequences, which has difficulties in handling very long  behavior sequences. 2). Sequences with the same length can be quite different in terms of time that carries completely different meanings. In this paper, we propose a Deep Interest Network with Sequence and Time Alignment (DINSTA) to solve the existing problems. Specifically, in the first step, a key-vector method was employed to extract behavior features in a non-sequential way, which preserves lifelong sequential information into a sparse representation. Next, we proposed a time-ware attention to calculate the correlation between behaviors by considering the influence of different time. Moreover, a Sequence and Time Alignment module is introduced to learn user interest representation by splitting the behavior sequence into custom designed time buckets. Finally, interest features, target item features, and user profile are fed to a deep neural network for predicting CTR. Experiments are conducted on two public datasets: one is an advertising dataset and the other is a production recommender dataset. Our proposed DINSTA outperforms other state-of-the-art  methods on both datasets.
Click-Through Rate (CTR) prediction plays an important role in many industrial applications,  and recently a lot of attention is paid to the deep interest models which use attention mechanism to capture user interests from historical behaviors.
However, most current models are based on sequential models which truncate the  behavior sequences by a fixed length, thus have difficulties in handling very long  behavior sequences.
Another big problem is that sequences with the same length can be quite different in terms of time, carrying completely different meanings.
In this paper, we propose a non-sequential approach to tackle the above problems. Specifically, we first represent the behavior data in a sparse  key-vector format, where the vector contains rich behavior info such as time, count and category.
Next, we enhance the Deep Interest Network to take such rich information into account by a novel attention network.
The sparse representation makes it practical to handle large scale  long behavior sequences.
Finally, we introduce a multidimensional partition framework to mine behavior interactions.
The framework can partition data into custom designed time buckets to capture the interactions among information aggregated in different time buckets. Similarly, it can also partition the data into different categories and capture the interactions among them.
Experiments are conducted on two public datasets: one is an advertising dataset and the other is a production recommender dataset. Our models outperform other state-of-the-art models on both datasets.

\end{abstract}

%%
%% The code below is generated by the tool at http://dl.acm.org/ccs.cfm.
%% Please copy and paste the code instead of the example below.
%%
% \begin{CCSXML}
% <ccs2012>
%  <concept>
%   <concept_id>10010520.10010553.10010562</concept_id>
%   <concept_desc>Computer systems organization~Embedded systems</concept_desc>
%   <concept_significance>500</concept_significance>
%  </concept>
%  <concept>
%   <concept_id>10010520.10010575.10010755</concept_id>
%   <concept_desc>Computer systems organization~Redundancy</concept_desc>
%   <concept_significance>300</concept_significance>
%  </concept>
%  <concept>
%   <concept_id>10010520.10010553.10010554</concept_id>
%   <concept_desc>Computer systems organization~Robotics</concept_desc>
%   <concept_significance>100</concept_significance>
%  </concept>
%  <concept>
%   <concept_id>10003033.10003083.10003095</concept_id>
%   <concept_desc>Networks~Network reliability</concept_desc>
%   <concept_significance>100</concept_significance>
%  </concept>
% </ccs2012>
% \end{CCSXML}

\ccsdesc[500]{Information systems~Recommender System}
% \ccsdesc[300]{Information systems~Redundancy}
\ccsdesc[500]{Information systems~Online Advertising}
% \ccsdesc[100]{Networks~Network reliability}

%%
%% Keywords. The author(s) should pick words that accurately describe
%% the work being presented. Separate the keywords with commas.
\keywords{Non-sequential Approach, User Interest Model, Time-aware Attention, Click-Through Rate, Sparse Representation}

%% A "teaser" image appears between the author and affiliation
%% information and the body of the document, and typically spans the
%% page.
% \begin{teaserfigure}
%   \includegraphics[width=\textwidth]{sampleteaser}
%   \caption{Seattle Mariners at Spring Training, 2010.}
%   \Description{Enjoying the baseball game from the third-base
%   seats. Ichiro Suzuki preparing to bat.}
%   \label{fig:teaser}
% \end{teaserfigure}

%%
%% This command processes the author and affiliation and title
%% information and builds the first part of the formatted document.
\maketitle

\section{Introduction}

Click-Through Rate (CTR) prediction is a core task in e-commerce such as online advertising and recommendation system, because it is directly related to revenues of the whole platform and also influences user experience and satisfaction.  For the CTR prediction model, user behavior modeling is a key part to improve the performance.

Recently,  many deep interest models  \cite{din,dien,dsin,bst,sim,hdin}
proposed for mining the user's historical behavior have been very effective for practical problems and also have been applied to the real online system successfully.  The basic idea behind these models is that user interests are diverse and it will be noisy if captured from user behavior data directly. So  attention mechanisms are used to activate historical behaviors with respect to the target item  \cite{din}. The behavior closely related to the target item will be given a large attention score while other behaviors will be filtered out by small weights.

Though the interest model is effective to improve performance, most models such as DIEN \cite{dien} and DSIN \cite{dsin} deal with historical behavior data in a sequential manner using GRU or LSTM.  A big limitation is that these sequential models can't handle very long sequences in practice.
%Recently, \cite{pi2019practice} reports successful handling of real data set where sequence can be as long as 1000. It is through the help of engineering and the co-design of the machine learning algorithm and online serving system.  \cite{sim} introduces cascaded search units so that max sequence length can reach up to 54000. This is through the system design and filtering long sequences into short ones.
Currently, some practical works are reported, such as the method in \cite{pi2019practice} is through the help of engineering and the co-design of the machine learning algorithm and online serving system, and the method in \cite{sim} is make use of the cascaded search technology to filter the long sequences into short ones.
To the best of our knowledge, there are no models specifically designed to deal with long-term behavioral data.

\begin{figure}[ht]
\begin{center}
\hspace{-0.3cm}
\includegraphics[width=0.51\textwidth,height=0.118\textwidth]{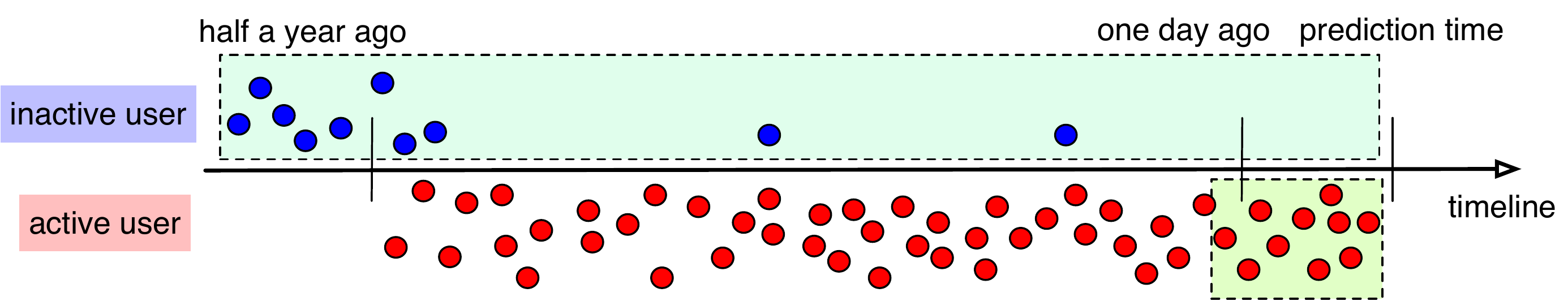}
\caption{ The behavior sequences of active users truncated by a fixed-length are quite different from that of inactive users.}
\label{fig:eg}
\end{center}
\end{figure}

We argue that sequential modeling on raw sequences may not be appropriate for long term behavior data in the case of CTR prediction.
On the one hand, sequential modeling techniques such as LSTM have difficulties dealing with long sequence and performance gets poor for sequence longer than several hundred in practice \cite{pi2019practice}. It is also well known that computation needed by recurrent network such as RNN, LSTM or GRU either in training or inference is not hardware friendly \cite{atrank}.  In addition, since fixed length input is required for sequential modeling in practice, a max length is usually used.  Thus sequences longer than the max length have to be truncated and shorter ones need padding.  This is not efficient in terms of data storage and computing.
On the other hand, different from sequence data in NLP where interval between words has little information, the time interval between user behaviors may vary greatly and carries valuable information.  Sequences with the same length can be quite different in terms of time, carrying completely different meanings, as shown in Figure \ref{fig:eg}. Intuitively, the behavior sequence patterns of active users are quite different from that of inactive users. For behaviors in a long time horizon, the detailed sequences happened long time ago may not be needed, while capturing the trend change at proper time scale is more important.

Motivated by the above observations, we propose a non-sequential approach to deal with the raw behavior data. We first conduct statistical processing on the original behavior data and summarize the count and time distribution of different behaviors. Then the aggregated information is expressed in the sparse key-vector format to eliminate redundancy of storage caused by the repetitions of the same behaviors. Based on the sparse representation, we enhance the Deep Interest  Network by introducing the time and count info into the attention calculation. Note that the sparse representation can also reduce the amount of data computation,  since the computation amount is related to the average number of keys in the samples, while in the sequence representation, it is determined by the fixed max number of keys in the samples.

%Motivated by the above observations, we propose a non-sequential approach to deal with the raw behavior data.  We first conduct statistical processing on the original behavior data and summarize the count and time distribution of different behaviors. The aggregated information is expressed in the sparse key-vector format.   Since the computation amount for sparse representation is related to the average number of keys in the samples, while in the sequence representation, the computation amount is determined by the fixed max number of keys in the samples,  the amount of data storage and computation for sparse representation can be greatly reduced. We also enhance the Deep Interest  Network (DIN) by introducing the time and count info into the attention  calculation.

For long term user behaviors, temporal trend contains valuable information, and it is important for the model to be able to mine such info.  Though our approach represents data in a non-sequential way, the time info is saved in the vector for each key. In order to mine valuable trend information or interactions among behaviors effectively, we propose a framework called Deep Interest Network with Multidimensional Partitions.  The behaviors are partitioned into custom designed time buckets.  Behaviors in the same time bucket are aggregated,  resulting in a much shorter sequence of aggregated information.  The sequence position after partition represents a fixed time bucket, and it is consistent among different samples.  The interaction layer such as self-attention is used to capture interactions among these time buckets. For long-term historical behavior data, it can effectively capture the dynamic interest changes of users over time.   Similarly, the framework also can partition the data into different categories and capture interactions among these categories if the categories are available.

The main contributions of this paper are summarized as follows:
\begin{itemize}
    \item We propose a non-sequential method by representing behavior data in the sparse key-vector format. The repetitions in the original behavior data can be aggregated as count and saved in the vector, thus long sequences can be compressed shorter.  Using sparse representation, the amount of data that our method can handle is determined by the average length of data, not the maximum length.  This makes our model able to deal with a much longer sequence under the same amount of data storage. % and computing.

    \item  We proposed a novel Time-aware Attention Mechanism which adds side information such as time and count into  the attention network.  Intuitively, the behaviors that happened several months ago have a very different impact on the user's interest than those of yesterday, and we should give them different weights. Temporal information is very important for mining historical behaviors spanning a long time horizon.

    \item We propose a Deep Interest Network with Multidimensional Partitions (DINMP).  The long behavior sequences are partitioned into custom designed time buckets and information within the same bucket is aggregated, resulting in the fixed length sequences of aggregated information.  The interactions among these time buckets are captured through self-attention, thus long-term temporal trend can be modeled. Similarly, the behaviors are partitioned into different categories, and interactions among categories can be captured.

    \item We conduct extensive experiments on two public datasets. On the Alimama's advertising dataset,  we show that our model can handle users' original behavior sequence as long as tens of thousands.  The testing results verify the effectiveness of our proposed DINMP.

\end{itemize}

Solutions discussed here are also applicable in many related tasks, such as conversion rate prediction and user preference modeling.

\section{Related Work}

Deep learning based methods have achieved great success in CTR prediction task \cite{widedeep,pnn,deepintent,Dual-fm}. %deepintent,Dual-fm,
 In early age, models such as DeepFM\cite{deepfm}, XDeepFM\cite{xdeepfm}, DCN\cite{dcn} use a deep neural network to capture interactions between features from different fields so that engineers could get rid of boring feature engineering works.
Recently, many  works focus on learning the users representation from historical behaviors, using different neural network architecture such as CNN \cite{cnn-seq,CosRec}, RNN\cite{rnn-session}, Transformer\cite{SSE-PT,dsin,BERT4Rec,atrank}  and Capsule\cite{MultiInterestNetwork}, etc.
Among them, the user interest models \cite{din,dien,dsin,hdin} which use attention mechanism to capture user interests from historical behaviors have been widely proven effective.

DIN \cite{din} first introduces the mechanism of attention to activate the historical behaviors w.r.t. given target item, and captures the diverse user interests successfully.
ATRANK\cite{atrank} proposes an attention-based framework modeling the influence between users' heterogeneous behaviors.
DIEN\cite{dien} points out that the temporal relationship between historical behaviors matters for modeling users’ drifting interest. An interest extraction layer based on GRU with auxiliary loss is designed in DIEN.
DSIN \cite{dsin}  assume the behavior sequences are composed of sessions and use self-attention to extract users’ interests in each session and apply Bi-LSTM to capture users' cross-session interests.
DHAN \cite{hdin} propose a hierarchical attention network to model users' interests from a higher-level attributes (e.g. category, price or brand) and progressively to lower-level attributes or items.

Due to the limitation of the sequential model, the above user interest models still have difficulties in handling very long behavior sequences.  A  practical system is proposed in \cite{pi2019practice} to handle long sequences with  the co-design of  the learning algorithm and online serving system for CTR prediction task.  Cascaded search units are introduced in \cite{sim} to further improve the max sequence length handled in the real system.

\section{DINMP}

In this section,  our work is introduced in two steps.
In subsection \ref{sect:edin}, we propose a sparse key-vector format for more efficient data representation,  and show that the Deep Interest Network can be implemented using this sparse non-sequential data representation.  Then traditional attention network in DIN is enhanced to take into account extra info such as time and count in the vector.
In subsection \ref{sect:edin-mp},  we propose a DIN with Multidimensional Partition (DINMP) framework to capture the interest evolution over time and the interaction of behaviors among different categories.
The overview of our model is shown in Figure \ref{fig:model}.

\subsection{Enhanced Deep Interest Network} \label{sect:edin}

\subsubsection{Deep Interest Network}\label{sect:sp_din}

The overall structure of DIN and its subsequent interest models \cite{dien,dsin} can be clearly described as follows:
\begin{enumerate}
    \item  \textbf{Feature Representation}. Overall, three groups of features are used as model inputs: user behavior,  ad features, other features e.g. user profile, and context features.
    \item \textbf{Embedding}. Embedding is a common technique. Generally the raw categorical features such as user behavior or ad will be transformed into low-dimensional dense vectors through embedding.
    \item \textbf{Sub-Network to capture Users' interests}. A core network is built from ad embedding and user behaviors' embedding. Its output vector stands for users' interest to the target ad.
    \item \textbf{Multiple Layer Perceptron(MLP)}. User interest vector, together with the embedding of ad features and other features are concatenated, flattened, and then fed into MLP. The softmax function is used at last to predict the probability of the user clicking on the target ad.
\end{enumerate}

The main idea of basic DIN is that the user's interest to the target ad is closely related to the part of behaviors that is highly similar to the target ad. Specifically, the model constructs an attention score to measure how closely the target ad is tied to each behavior of the user firstly and then the embedding of user's historical behaviors are aggregated by weighted sum pooling operation, that is,
\begin{align}\label{eq:din}
\varepsilon_i = \sum_{j=1}^{N_i} a(e_j, q_i) e_j
\end{align}
where $q_i$ is the embedding vector of ad in the current sample $i$, $\{e_j: j =1,2 \cdots, N_i\}$ are the embedding vectors of user's historical behaviors, and  $a(e_j, q_i)$ denote the attention function. The formula (\ref{eq:din})  can also be understood as that we assign the weight $w_{ij}$  to the embedding vector $e_j$ of behavior $j$ in the current sample $i$.

Though the attention and sum pooling operation in (\ref{eq:din}) does not require sequential relations, in practice the DIN as well as other interest methods take the behavior sequence as inputs.
 In the model implementation,  the sequence data is usually represented  by a dense tensor with fixed dimension.  A max length is picked, and every sequence needs to be either truncated or padded according to this max length. This causes a lot of waste both in storage and computation for short behavior sequences.

 There is another waste related to the repetition of the same behavior in the user's historical behaviors.
Taking user transaction data as an example, the user may have repetitive browsing behaviors for the same product at different times. In the current  interest models,  such repetitive behaviors  will be included many times in the raw sequence data,  which causes the length of the sequences  to be extremely large.

\begin{figure}[ht]
\begin{center}
\hspace{-0.3cm}
\includegraphics[width=0.46\textwidth,height=0.3\textwidth]{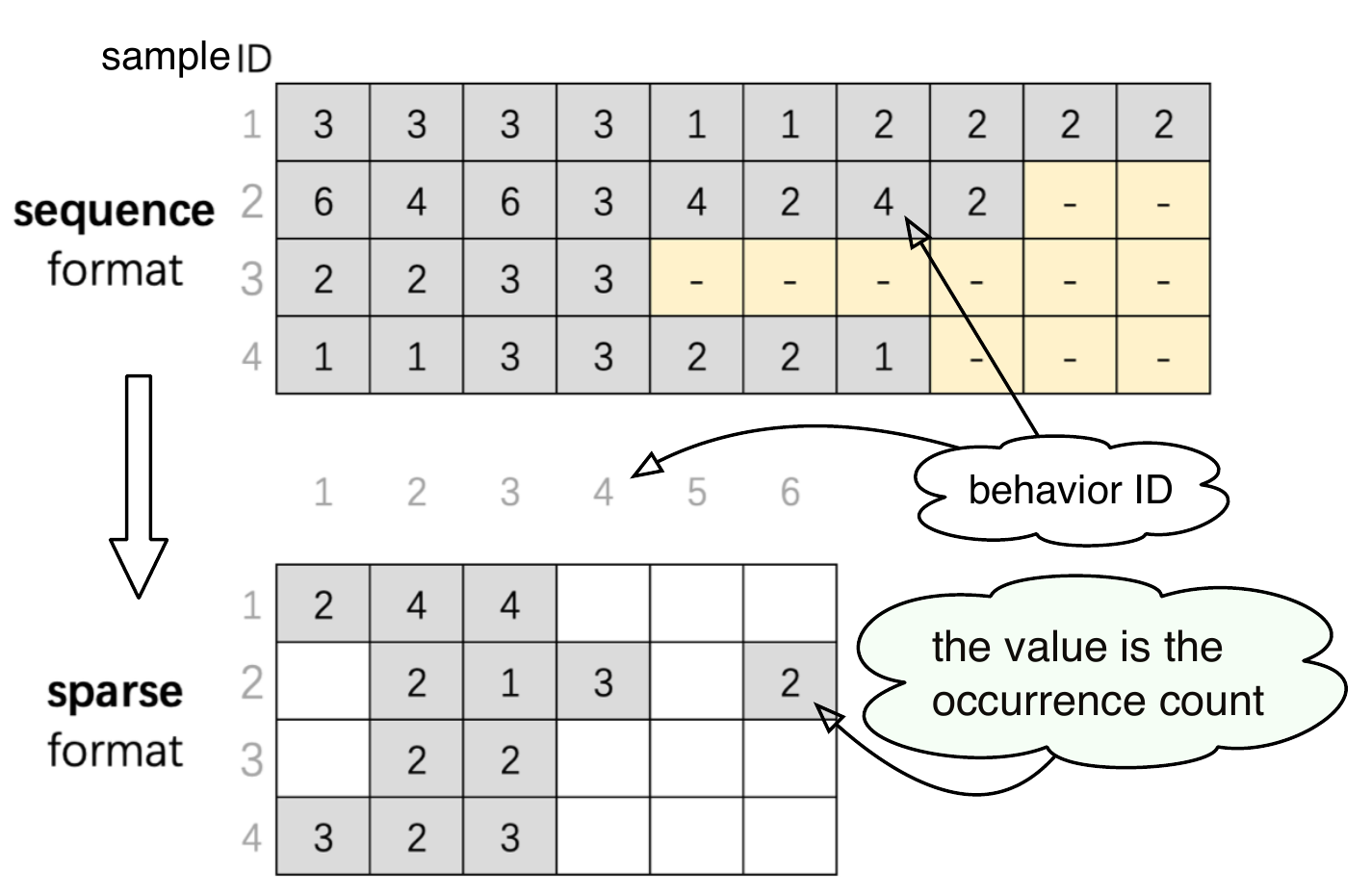}
\caption{An example is given to illustrate the differences in data representation between the sequence format and  the sparse format.  The latter is more compact due to no duplicates and padding. Notice that the sparse matrix only  stores the non-zero elements.}
\label{fig:kv}
\end{center}
\end{figure}

 Base on these observations, we introduce a non-sequential sparse key-vector format in order to represent long behaviors more efficiently. The key idea is shown in  Figure \ref{fig:kv}.

\begin{figure*}[ht]
\begin{center}
\hspace{-0.3cm}
\includegraphics[width=1.0\textwidth,height=0.45\textwidth]{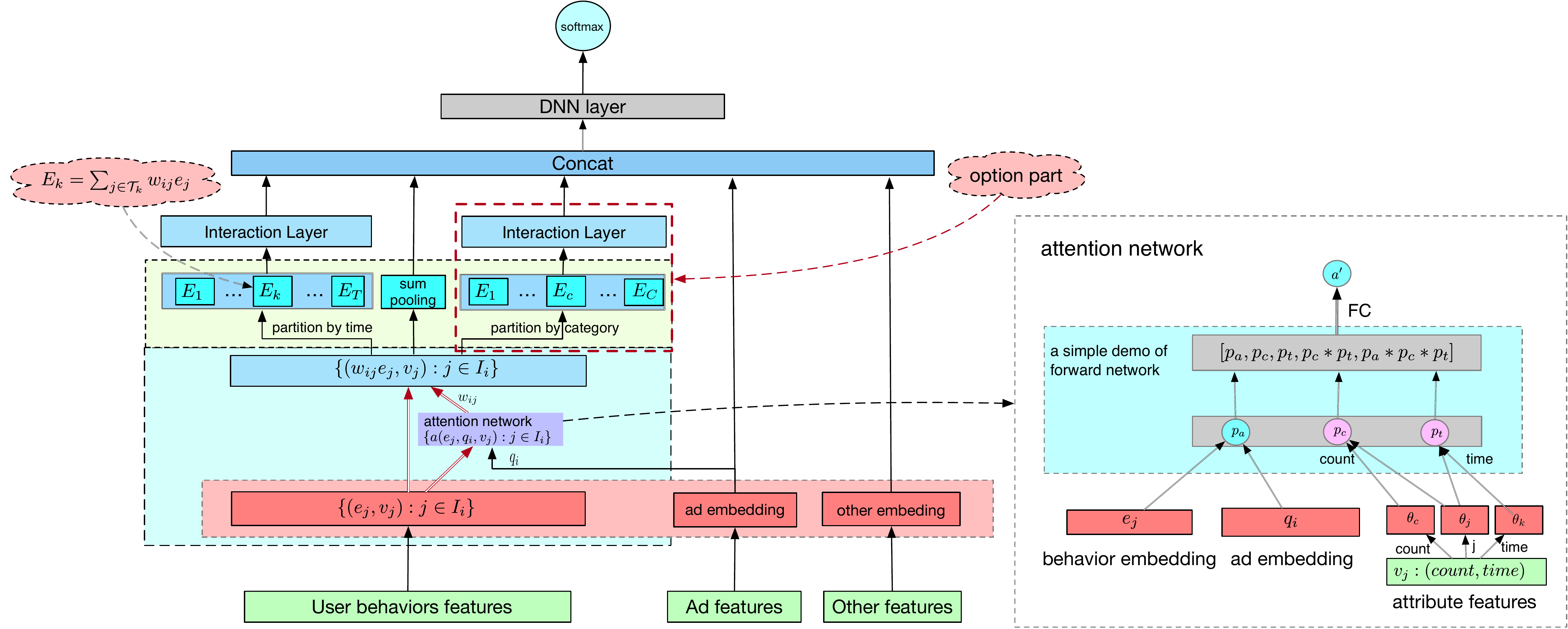}
\caption{The overview of our proposed model EDINMP.  The behavior embeddings are partitioned by time windows and categories to capture the interest evolution and the interactions among different categories.  All the operations are based on the sparse key-vector representation. }
\label{fig:model}
\end{center}
\end{figure*}

\subsubsection{Sparse Key-vector Representation for Behavior Data} \label{subsect:kv}
The details of the non-sequential format are introduced next.
Note that the user's raw historical behaviors may have repetitive behaviors at different times.
We consider a mild aggregation of the original data based on the statistics of each behavior to eliminate redundancy. We summarize the user's behaviors and rewrite the behaviors of user $i$ as
\begin{equation} \label{eq:key-vector}
\{ (j, v_j ):  j \in J_i\},
\end{equation}
where $v_j$ is a vector of summary info. Some specific forms are as follows:
\begin{itemize}
   \item[a).] $v_j = [c_{j} ]$ that counts the occurrence times of behavior $j$ directly.  We emphasize that this format is sufficient for the basic DIN model.
   % \item[a).] $v_j = [c_{j_1}, \cdots, c_{j_k} ]$. The numbers of repeated occurrences at different time buckets. We emphasize that  the original information is not lost in this form.
    \item[b).] $v_j = [t_j, c_j]$ that consists of the time of the last occurrence and  the number of occurrences.  Note that the time info is a piece of side information which is not included by the current interest models.    % It loses some temporal information, but can greatly reduce the amount of raw data.
    \item[c).] $v_i = [t_i, c_i,  \alpha_i]$. If the behaviors have additional category partitions, the vector can be set as this format, where $\alpha_i$ represents the category of $i$-th behavior belonged.
\end{itemize}

We call (\ref{eq:key-vector}) as the  key-vector format for behavior data representation which is an extension of  the key-value format in machine learning.
Similar to the sparse matrix for data representation in key-value format, we use triplet $(I,J, V)$
for behaviors in key-vector format, where $I, J$ stands for the row and column index vectors, respectively and
$V$ is a matrix  consisting the vectors of all observed elements. Without duplicates or padding,  this compact way is more suitable for DIN model.

We emphasize here that the key-vector format reduces the amount of behavior data dramatically in the real world dataset, and it enables to handle long-term behavior data.
See Table \ref{tab:data-stat}.

\subsubsection{ Enhanced Deep Interest Network} \label{subsect:attention}

Due to the flexibility of the key-vector format,  a lot of additional information of behaviors such as the time, count and category can be added to the attention score. Notice that  there is a lot of side information that is not considered in the interest models.  For example, the time information is ignored in DIN\cite{din}, DIEN\cite{dien}, DSIN\cite{dsin}.
We propose a novel Time-aware Attention Mechanism to take the additional information into account.  The original attention network  $a(e_j, q_i)$ in (\ref{eq:din}) is substituted by a new one $a'(e_j, q_i, v_j)$.
 \begin{align}\label{eq:sdin}
\varepsilon_i &= \sum_{j \in I_i} a'(e_j, q_i, v_j) e_j
              = \sum_{j \in I_i} a'(e_j, q_i, t_j, c_j,  \alpha_j) e_j
\end{align}
here $v_j= (t_j, c_j, \alpha_j)$ means the last occurrence time, the count and the category  of behavior $j$.
After the modification of (\ref{eq:sdin}), we denoted the non-sequential version of DIN as the Enhanced DIN or EDIN for short.
EDIN is compatible with the DIN model while introducing additional  information as an extended version.

In the rest of this subsection, we will describe how to construct the new attention network in details.
We hope that the extra information will be incorporated into the attention network to improve the performance by designing a more reasonable network structure.

Though there is no definitive way on  how to design the new attention network,  we want to keep a few intuitive principles:
\begin{itemize}
    \item We keep the old attention as a part, which takes the embedding of target ad and behavior as inputs.
    %\item For time information, an empirical assumption is that if a behavior happened earlier, it would have less impact on the outcome.
    \item The impact of additional info such as time and count should be personalized for each behavior. We notice that some behaviors occur frequently while others are always inactive but important.
\end{itemize}

Motivated by the above principles and through many testing, we design a time-aware attention network as follows.

First, the count and time are designed to interact with the behavior.  Take the time $t_j$ for example, it is parameterized as
\begin{align}\label{eq:timefactor}
p_t &=  e_i^T \theta_k.
\end{align}
%$$p_t = \theta(j, k):= \theta_j^T \theta_k.$$
where $\theta_j$ is the parameter vector of behavior $j$ and $\theta_k$ is the parameter vector for time $t_j$  assigned into the $k$-th time bucket.
Similarly, the count is parameterized as $p_c = \theta_j^T \theta_c$ and the category is parameterized as $p_\alpha = \theta_j^T \theta_\alpha$.
Note that the new vector $\theta_j$ for behavior $j$ is introduced in addition to the embedding $e_j$ or simply transformed from it $\theta_j = \theta(e_j)$.
The dimension of $\theta_j$ is small compared to the embedding vector $e_j$.

Beside  the old attention value $p_a = a(e_j, q_i)$, we have three other factors $p_t, p_c, p_\alpha$ at hand and the final output is built by a nonlinear combinations:
\begin{align*}\label{eq:att}
  a' = w_0  p_a +  w_1  p_t +  w_2 p_c + w_3  p_\alpha + w_4  p_t p_c p_\alpha + w_5  p_a  p_t p_c p_\alpha
\end{align*}
where $w_i$ are the parameters of the network.  The time-aware attention network is shown in Figure \ref{fig:model}.

\subsection{DIN with Multidimensional Partition} \label{sect:edin-mp}

Similar to DIN,  the Enhanced DIN  calculates the users' interest for each behavior separately, but lacks the ability to mine interactions among behaviors.  To overcome this problem, we propose a DIN with Multidimensional Partition (DINMP), where the behaviors can be partitioned by the time windows or by the behavior categories respectively.

\subsubsection{Multidimensional Partition}

Currently, user interest models do not take time into account when organizing behavior sequences, although some works attempt to use time information,  e.g., \cite{atrank} includes the time information in a simple way i.e. adds the embedding of time to the embedding of behavior directly.
Modeling Behavior sequences without considering temporal information can cause problems. For example in Figure \ref{fig:eg}, if the behavior sequences are truncated by a fixed length, the behavior time range of active users may be quite different from that of inactive users.
Most of the behaviors of inactive users may have happened half a year ago, while the behaviors of active users have happened just one day ago. For such an example, the two sequences have very different meanings, but such difference can not be captured by existing sequential models because valuable time info is mostly ignored.

%\begin{figure}[ht]
%\begin{center}
%\hspace{-0.23cm}
%\includegraphics[width=0.49\textwidth,height=0.118\textwidth]{inactive-user-eps-converted-to.pdf}
%\caption{ The behavior sequences of active users truncated by a fixed length are quite different from that of inactive users.}
%\label{fig:eg}
%\end{center}
%\end{figure}

%The assumption in DSIN model \cite{dsin} is that the sequences of behaviors are composed of sessions, where sessions are user behaviors separated by their occurring time and divided in the principle of whenever there exists a time gap of more than 30 minutes.
%Intuitively, DSIN takes sessions into account making it more capable and reasonable.
%However, this method also ignores the difference in the time range of user behavior, especially fails to effectively solve the problem of modeling long-term behaviors. Moreover, extracting new sequences with session flags from raw data is not a easy task for data engineers. Thus, it is not a very common approach in dealing with real-world business problems.

Motivated by the above insights, we align all behaviors of the user before the prediction on the timeline, and we divide it according to the natural date. For example, the behaviors in recent one year can be partitioned by a custom-designed set of time points such as 1 day ago, 2 days ago, 4 days ago, 1 week ago, 2 weeks ago, 1 month ago, 2 months ago, 3 months ago, 6 months ago, etc. After partition, all the behaviors are aligned into fixed groups.

%For example, on our first experimental dataset, we use natural dates to divide time windows by days as follows:
%$$[-21, -14), [-14, -7), [-7, -3), [-3, -1), [-1,0)
%$$
%where [-1,0] represents the time period from 0 a.m. of the current day to the current time and other intervals have similar meanings. After division, all the behaviors are aligned into 5 groups.

For all behaviors within each time window, we can use the module described in EDIN (\ref{sect:sp_din}) to capture users' interest. EDIN represents user behaviors in the key-vector format as inputs, and returns an aggregated vector  $E_k$ as user' interest in that time window.
Finally, all of the outputs are collected in sequence as $[E_1, \cdots, E_k, \cdots, E_T]$.
Notice that it forms a new time-aligned sequence of aggregated user behavior representations,
which allows the neural network to add some higher-order layers,
such as transformer or concatenation,  to improve the model's performance further.

The above partition and aggregation way to improve performance can be extended to multiple dimensions. For example, a natural way is to partition and aggregate by category of the behaviors.

\subsubsection{Model Overview}
The core network structures of our DINMP are shown in Figure \ref{fig:model}. Most key calculations are listed as the following:

% \bigskip
\begin{center}
\fbox{
\parbox{8.3cm}{
  \noindent {\bf Input}: User behavior feature in key-vector format, or mathematically as follows $\{ (j, v_j): j \in I_i\},$ where $I_i$ is the index set of keys in sample $i$.
 \begin{enumerate}
    \item [  1.] Look up embeddings for behaviors, and we get $\{ (e_j,v_j): j \in I_i\}$.
    \item [ 2.]  Use attention network to compute weight between ad and behavior i.e. $w_{ij} = a'(e_j, q_i, v_j)$, here $q_i$  means ad embedding.
    \item [  3.] Compute weighted embedding sets, and get a new set $\{ (w_{ij}e_j,v_j): j \in I_i\}$.
    \item [4.]  Partition the behaviors and aggregate embeddings.
    \begin{enumerate}
        \item   Denote  $\{{\cal T}_k\}_{k=1}^T$  be the partition by  time window,  i.e.,  $I_i = \cup_{k=1}^T  {\cal T}_k$. Aggregate the embeddings as \\
              $E_k = \sum_{j \in {\cal T}_k} w_{ij} e_j, \quad k= 1, \cdots, T.$
       \item   Let  $\{{\cal C}_k\}_{k=1}^C$  be the partition of $I_i$ by  category,  \\ i.e.,   $ I_i  =  \cup_{k=1}^C  {\cal C}_k $ .  Aggregate the embeddings as \\
              $E_c = \sum_{j \in {\cal C}_c} w_{ij} e_j, \quad c = 1, \cdots, C.$
       \item Sum pooling i.e., aggregate all the weighted embeddings.
              $E = \sum_{j \in I_{i}} w_{ij} e_j $
    \end{enumerate}
\end{enumerate}
}
}
\end{center}

Notice that the network described in the above box can be implemented with TensorFlow \cite{abadi2016tensorflow} using sparse representation methods.

\section{Experiments}
In this section, we  mainly present our experiments in detail, including three datasets introduction, evaluation metric, experimental setup, comparison with baseline methods,  ablation study,  and visualization analysis. Experiments on public datasets with user behaviors demonstrate the effectiveness of  our proposed approach that outperforms state-of-the-art methods on the CTR prediction task. Both the public datasets and experiment codes are made available.

\subsection{Datasets and Experimental Setup}

{\bf Advertising dataset}: Advertising dataset is a public dataset released by Alimama \footnote{https://tianchi.aliyun.com/dataset/dataDetail?dataId=56}, an online advertising platform in China. It contains 26 million records from ad display/click logs of 1 million users and 800 thousand ads in 8 days.
We used the first 7 days’s logs as training samples, and the last day’s logs as test samples. The user behavior data set covers  the shopping behavior in 22 days.  We process the user behaviors data by only extract the {\bf ipv}(click) behaviors and ignore other three types ({\bf buy, cart, fav}) of behaviors as in \cite{dsin}.

Sample information is given in the following table \ref{tab:data-stat}. Note that after sparse processing, the average number of keys in each sample is only 175. In the detailed sequence, the maximum length is over 40 thousands. It is nearly 250 times as long as processed.

 \begin{table}[!htbp]
  \centering
    \caption{ The description of  the advertising dataset.  By simple aggregation, the raw behavior sequence data can be reduced into a very compact key-vector format.}
  \begin{tabular}{|c|c|c|c|}
   \hline
   train sample  &	$2.37 \times 10^7$ &	test sample	& $2.8 \times 10^6$ \\ \hline
 max \#behavior &	43251	&avg \#behavior	&606  \\ \thickhline
 max \#key& 	3809	&avg \#key&175  \\ \hline
 \#key  $<$ 500	& 96\%  &  \#key $<$ 400   &	92\%  \\
     \hline
\end{tabular}
\label{tab:data-stat}
 \end{table}

\textbf{Taobao}\footnote{https://tianchi.aliyun.com/dataset/dataDetail?dataId=649}:
Taobao Dataset is a collection of user behaviors from Taobao’s recommender system\cite{treeDM}, which consists of different types of user behaviors such as click, purchase, etc.  It contains user behavior sequences of about one million users. We take the click behaviors for each user and sort them according to time in an attempt to construct the behavior sequence. Assuming there are $T$ behaviors of a user, we use the former $T-1$ clicked products as features to predict whether users will click the $T$-th product. Note that our model takes all the $T-1$ clicked products as input features, while for other comparison methods,  the behavior sequence is truncated at length 200.

Sample information is given in the following table \ref{tab:data-tb}.
the average number of behaviors is 90, which is small compare to truncated length 200.
 \begin{table}[!htbp]
  \centering
    \caption{ The description of  the Taobao recommender dataset. The average number of behaviors is 90.}
  \begin{tabular}{|c|c|c|c|}
   \hline
   train sample  &	$691456$ &	test sample	& $296192$ \\ \hline
 max \#behavior &	846	&avg \#behavior	&90  \\ \hline
      \#behavior  $<$ 400	& 99\%  &  \#behavior $<$ 200   &	90\%  \\
      \hline
  \end{tabular}

\label{tab:data-tb}
 \end{table}

\textbf{Amazon(Electro)}: Amazon Electronics data set is used to evaluate our proposed multi-dimensional partition module. The entire dataset contains product review information and metadata from Amazon. Our task is to predict $(k+1)$-th reviewed item by using the first k($\geq 5$) reviewed items. We construct one positive sample by taking user’s historical reviews, and generate three negative samples from items not reviewed by this user. The numbers of training and testing sets are 697092 and 190468. Notice that this data set  includes coarse category information that is different from the above datasets and all the items can be categorized into 51 groups.

% 说明一下实验参数, 没写好
\textbf{Experimental Setup}: For all our tests, we use Adam\cite{Adam} solver and set the learning rate to be 0.001. Layers of DNN are set by $(128, 256, 80, 256)$. For taobao dataset, the behaviors are represented in two features, i.e.,  $(items: [category, time])$ and $(category,[time])$, and the dimension of embedding is 16 which is same as all the other methods \cite{pi2019practice}. For alimama dataset, two behavior features are $(brand: [count, time])$ and $(category: [count,time])$, and the dimension of embedding for behaviors features is 8 while that for other features is 4. We take AUC as the metric for measurement of model performance.

\subsection{Compared Methods}

The compared methods are listed here.
\begin{itemize}

    \item \textbf{Embedding\&MLP}. These methods represent a general paradigm in the early CTR prediction field. It first maps large scale sparse features  into low dimensional embedding space. Afterward, embedding vectors  are transformed into fixed-length vectors in a group-wise manner fed into multilayer perceptron (MLP) to predict the result.

    \item \textbf{YoutubetNet}\cite{ytube}. YoutubeNet is a technically designed model which treats users’ historical behaviors equally and utilizes average pooling operation to obtain user representation.

    \item \textbf{Wide\&Deep}\cite{widedeep}. Wide\&Deep is trained by wide linear models and deep neural networks jointly, which combines the benefits of memorization and generalization.

    \item \textbf{DIN}\cite{din}. A local activation unit is designed in DIN to adaptive learn the interest's representation of the user from historical behaviors. This representation vector varies over different ads, improving the expressive ability of the model greatly.

    \item \textbf{GRU4Rec}\cite{rnn-session}. It is the first work to exploit the RNN models in the recommender system  and introduces a new ranking loss function suited to the task of training these models.

    \item \textbf{RUM} \cite{sum}. RUM leverages the external memory matrix to store and update the user’s historical records, which also enhances the per item-level or feature-level correlations between  the user’s historical records and future interests.

    \item \textbf{DIEN} \cite{dien}.DIEN extracts latent temporal interests from user behaviors and models interests evolving process. Auxiliary loss makes hidden states more expressive to represent latent interests and AUGRU models the specific interest evolving processes for different target items.

   \item \textbf{DSIN} \cite{dsin}. It divides users' behavior sequences into sessions by their occurring time. It extracts users' interests in each session first and then model how users’ interests evolve and interact among sessions.

    \item \textbf{MIMN}. MIMN adds memory utilization regularization and memory induction unit in a traditional memory network that make it more efficient for dealing with long  user behavior sequences under limited storage and brings remarkable gain over model performance.

\end{itemize}

\subsection{Metrics}

Area Under ROC Curve (AUC) is a widely used metric in CTR prediction field, which means the probability that a randomly chosen positive example is ranked higher than a randomly chosen negative example, reflects the ranking ability of the model. Following the previous works,  AUC is regarded as the evaluation metric to validate the performance of the proposed DINMP for CTR prediction on the Alimama and Taobao dataset. The definition of AUC is described as follows:
$$AUC = \frac 1{m^+ m^-} \sum_{x^+ \in D^+} \sum_{x^- \in D^-}  {\bf 1} (  f(x^+) >  f(x^-)  ) $$
where $D^+$ is the collection of all positive examples, $D^-$ is the collection of all negative examples, $f(.)$ is the result of the model’s prediction of the sample $x$ and ${\bf 1}$ is the indicator function.

RelaImpr \cite{hdin} metric is used to measure the relative improvement over models, which is defined as:
$$
RelaImpr = \left( \frac { AUC(\text{messured model}) -0.5}{AUC(\text{base model}) - 0.5}  -1    \right) \times 100\%.
$$

\subsection{Results on Alimama Datasets}

% Table \ref{tab:madata}  describes the results of our model with DIN, DIEN and DSIN on public datasets.On the Alimama advertising dataset, the results of DIN, DIEN, DSIN are adopt from \cite{dsin} directly in which users'  behavior sequences are truncated in a fixed length with $N=200$ as reported while our models make use of all behaviors that cover 22 days and we emphasize that the average  number of keys is 175.  Here, DSIN is treated as the state-of-the-art baseline. Although DSIN beats  basic user interest model (DIN and DIEN) with its capability of finding behaviors' session information, EDIN and DINMP beat DSIN, since firstly our models capture the users' all behaviors, and secondly, our models make use of the behavior time as an important factor in attention score to achieve effective interest expression. The sparse representation network helps EDIN to beat DSIN with remarkable AUC improvement,  i.e.  4.1\% RelaImpr. The structure of grouping the keys' embedding by time windows help EDIN-T to get  4.9\% RelaImpr comparing to EDIN.
To evaluate the effectiveness of the proposed DINMP,  we compare DINMP to some state-of-the-art methods for CTR prediction. Existing methods used in our comparative experiments mainly include YoutubeNet, Wide\&Deep, DIN, DIEN,  as well as DSIN.

\begin{table}[h]
  \caption{Model performance (AUC) on Alimama}
  \label{tab:madata}
  \setlength{\tabcolsep}{4mm}{
    \begin{tabular}{lcc}
    \toprule

     Model & AUC & ReIaImpr (\%) \\
    \midrule
        YoutubeNet & 0.6313 & -1.28  \\
        Wide\&Deep & 0.6326 & -0.30  \\
        DIN & 0.6330 & 0.00\\
        DIEN & 0.6343 & 0.98\\
        DSIN & 0.6375 & 3.38 \\
        DINMP & 0.6442 & 8.42 \\
    \bottomrule
  \end{tabular}
  }
\end{table}

\textbf{Experimental results on the Alimama dataset}: In this part, the performance of compared baseline approaches are directly adopted from Feng et al. experiments \cite{dsin}. As described in Table \ref{tab:madata},  making comparisons among previous models, we report several conclusions: 1) As expected,  DIN has achieved superior performance compared with simple model including YoutubeNet, Wide\&Deep with improvements of 0.0027 and 0.0004 on AUC. It validates again the importance of user interest. 2) Moreover, DSIN is treated as the state-of-the-art baseline beating  basic user interest model (DIN and DIEN) with its capability of finding behaviors' session information. 3) The proposed DINMP method yields a new state-of-the-art performance for CTR prediction on the Alimama advertising dataset, which exceeds DSIN with improvements of 0.0067. It could be that our models make use of all behaviors covering 22 days whereas previous approaches  truncated behavior sequence in a fixed length resulting in information lose. Another reason is that DINMP adopts time-aware attention mechanism that makes use of the behavior time as an important factor to achieve effective interest expression.

\subsection{Result on Taobao Datasets}
In this section, we make an experiment on Taobao dataset to further verify the efficiency of the proposed DINMP model. Some state-of-the-art baselines for CTR prediction on this dataset are applied to the comparative experiments, e.g. RUM, DIEN, and  MIMN. Among them, the data are quoted from the experiment of Pi et al

\begin{table}[h]
  \caption{Model performance (AUC) on Taobao Datasets}
  \label{tab:atdata}
  \setlength{\tabcolsep}{4mm}{
    \begin{tabular}{lcc}
    \toprule
    % \multirow{}{}{Model} &  \multicolumn{2}{c}{Taobao}  &  \multicolumn{2}{c}{Amazon} \\
     Model & AUC & ReIaImpr (\%) \\
    \midrule
        Embedding\&MLP & 0.8709 & 3.24  \\
        DIN & 0.8833 & 0.00  \\
        GRU4REC & 0.9006 & 4.51 \\
        RUM & 0.9018  & 4.83 \\
        DIEN & 0.9081 & 6.47 \\
        MIMN & 0.9179 & 9.03 \\
        DINMP & 0.9342 & 13.28 \\
    \bottomrule
  \end{tabular}
  }
\end{table}

\textbf{Experimental results on the Taobao dataset}: As shown in Table \ref{tab:atdata}, we have the following observations.1) The baseline deep learning approach Embedding\&MLP is the lowest performance, which only obtains 0.8709 AUC. 2) By focusing on user interest information, DIN improves 0.0124 compared with Embedding\&MLP.  3) Owing to capture interest evolving process, DIEN has made a further promotion in performance. 4) Thanks to a novel memory-based architecture, MIMN achieves the best baseline performance by capturing user interests from long sequential behavior data. Although existing methods obtains great breakthrough, it still faces the limited sequence length problem. 5) Compared with all approaches, the proposed DINMP contributes a new state-of-the-art (0.9342 AUC) by employing non-sequential behavior representation.

\begin{table*}
  \caption{Ablation Studies Result on Alimama Advertising Dataset and Taobao Recommender Dataset. }
  \label{tab:atdata}
  \setlength{\tabcolsep}{6mm}{
    \begin{tabular}{lcccc}
    \toprule
    \multirow{2}{*}{Model} &  \multicolumn{2}{c}{Alimama Advertising Dataset} & \multicolumn{2}{c}{Taobao Recommender Dataset}  \\
      & AUC & ReIaImpr (\%) & AUC & ReIaImpr (\%)  \\
    \midrule
    	DIN  & 0.6330  & 0.00  & 0.8833 & 0.00 \\
        DINSKV 			& 0.6424  & 7.07       & 0.8876 & 1.12 \\
        EDIN 				& 0.6436  &  7.97     & 0.9285 & 11.79  \\
        DINMP    & 0.6442  & 8.42      & 0.9342 & 13.28 \\
    \bottomrule
  \end{tabular}
  }
\end{table*}

\subsection{Ablation Studies}
In this section, we perform ablation studies to verify the effectiveness of each module (i.e., Sparse Key Vector representation, the Time-aware Attention mechanism, Multi-dimensional Partition module) for the proposed method DINMP on three published datasets. In detail, we design five different models and each of them is described as follows:

\begin{itemize}

    \item \textbf{DIN}. In ablation studies, we employ basic DIN  as a baseline system, which takes truncated behavior sequences as input to obtain user interest representation.

    \item \textbf{DINSKV}.  Based on basic DIN, we implement a sparse version of DIN by representing behavior data in sparse key-vector format. Due to the sparse representation, the original behavior data is compressed shorter and the amount of data that the model can handle is determined by the average length of data, not the maximum length. Thus, it is able to tackle all behavior data as input.

    \item \textbf{EDIN}. Considering the impact of side information such as time on user interest, on the basis of DINSKV, we use a time-aware attention mechanism to enhance the local activation unit in DIN.

    \item \textbf{DINTP}. Adding time partition branch to EDIN model,  in this way,  DINTP can further capture the interactions among different time buckets.

    \item \textbf{DINMP}. DINMP makes further optimization on DINTP method by adding category partition, which captures the interactions among different category.

\end{itemize}

%  描述一下实验结果

\textbf{Ablation Studies on Alimama Dataset: }From the result in Table \ref{tab:atdata}, we have the  following observation on Alimama dataset. First, the model effect has been greatly improved on Alimama dataset when DIN use Sparse Key Vector (SKV) representation. Compared with baseline model DIN,  DINSKV  rise from 0.6330 to 0.6424 in AUC, which proves the effectiveness of SKV format. Here, we point out that since the SKV representation was effective, the average number of keys involved in each sample is only 175, and the data storage do not increase. Second, by adding a Time-aware Attention (TAA) mechanism in DINSKV, which takes time, count and category info into account, the effect on Alimama datasets has been improved 0.0012 AUC. It proofed that TAA has digged out  user interest by considering time evolution  in behavior. Finally, by introducting Multi-dimensional Partition (MP) module, DINMP makes a remarkable improvement (0.6442 vs. 0.6436) on performance compared to EDIN. It indicates that MP module brings an effective gain.

\textbf{Ablation Studies on Taobao Dataset: }In Taobao dataset, we obtain the same observation with Alimama dataset, which further proves the efficiency of our proposed methods. It is worth noting that AUC score becomes 0.9285, when adding TAA module in DINSKV rising 0.0409. Moreover,  the AUC increase from 0.9285 to 0.934 comparing DINMP with EDIN, which is a very significant lift. It shows tha MP module further improves the performance.

\begin{table}[h]
  \caption{Ablation Studies Result on Amazon Dataset}
  \label{tab:mpc}
  \setlength{\tabcolsep}{4mm}{
    \begin{tabular}{lcc}
    \toprule

     Method & AUC & ReIaImpr (\%)\\
    \midrule
        EDIN & 0.8210 & 0.00  \\
        DINTP & 0.8237 & 0.80 \\
        DINMP & 0.8285 & 2.30\\
    \bottomrule
  \end{tabular}
  }
\end{table}

\textbf{Ablation Studies on Amazon Dataset: }This part mainly verifies the effectiveness of the proposed multi-dimensional partition module because only Amazon dataset satisfied multi-dimension. As is shown above, DINTP outperforms EDIN (0.819 vs. 0.815) on AUC, which shows that the time partition of behavior sequence can lead to better performance. Moreover, compared with DINTP model which only exploits time partition, DINMP benefits from the combination of time and category partition improving 2.3\% on ReIaImpr, which indicates the superior of multi-dimensional partition.

\subsection{Visualization of Time-aware Attention}
%In this part, we further evaluate whether the proposed time-aware attention can learn more accurate user interest in CTR prediction task. The visualization result of time-aware attention is presented in Figure \ref{fig:ptime}, which portrays the change of weight value follow time.

\begin{figure}[ht]
\begin{center}
% \hspace{-0.3cm}
\includegraphics[width=0.46\textwidth,height=0.3\textwidth]{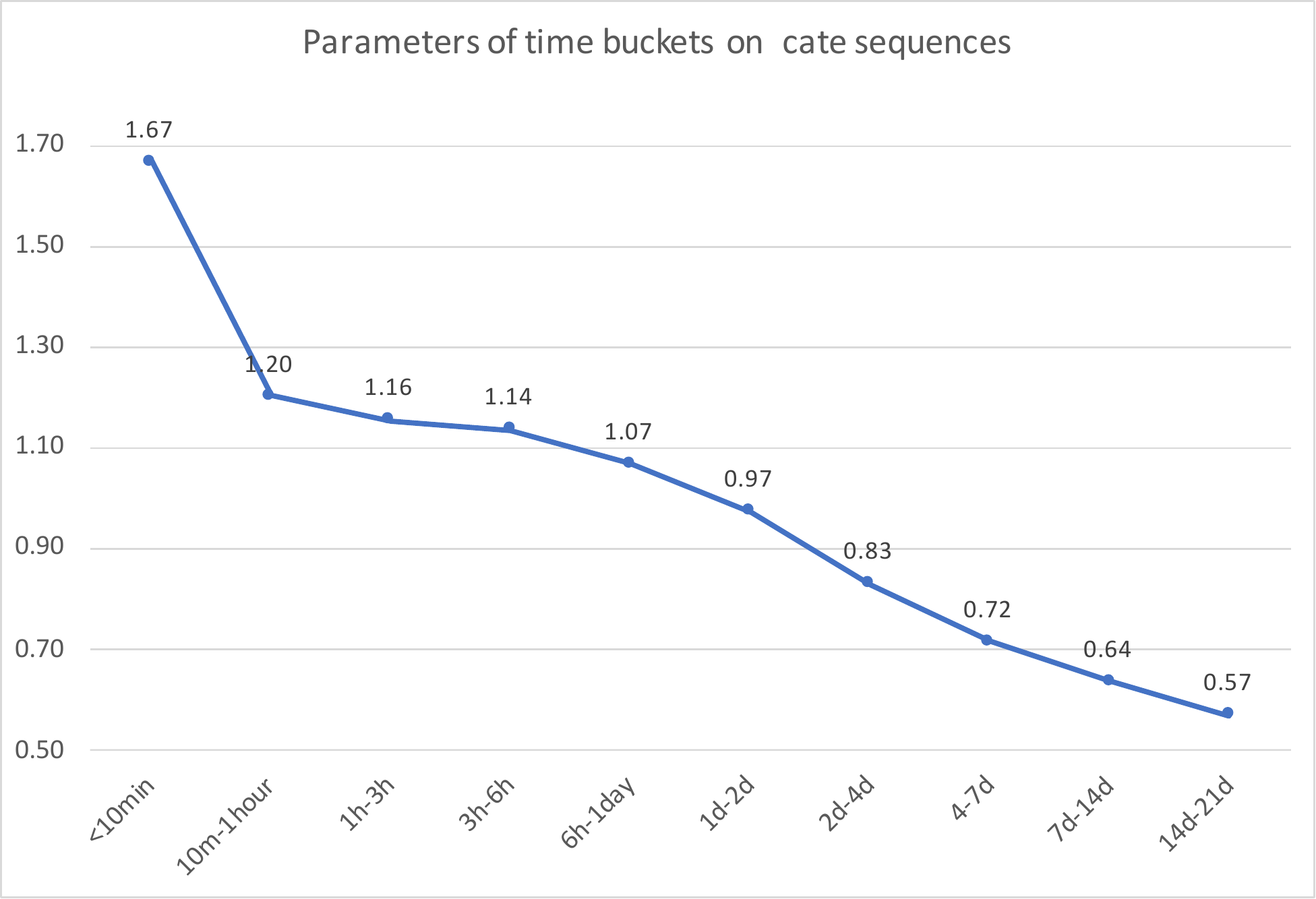}
\caption{The time factor $p_t$ over different time buckets on Alimama dataset.}
\label{fig:ptime}
\end{center}
\end{figure}

To further verify the superiority of the proposed time-aware attention mechanism, we perform a visualization experiment to visualize the time factor in the TAA mechanism, shown in Fig. \ref{fig:ptime}. For the sake of visualization, we simplify the time factor in Eq. \ref{eq:timefactor} to be $p_t = \theta_k$, where $\theta_k \in \mathbb{R}^1$ is parameter for time $t$ assigned into  the $k$-th time buckets. We retrained our DINMP model on the Alimama dataset and saved the final parameters.

We plot the time factor $p_t$ over different time buckets in Fig. \ref{fig:ptime} corresponding category behavior feature on the Alimama dataset.  From this figure, we can intuitively see that the time factor is almost getting smaller as time goes on. In particular, the time factor of the recent 10-minute bucket is nearly three times as large as that of the 2-3-week bucket. The above phenomenons indicate that time factor is discriminative, and our time-aware attention method with time and category information can effectively make the network focus on user recent interest.

\section{Conclusions}

In this paper, we propose  a Deep Interest Network with Multi-dimensional Partition (DINMP) to improve the performance of CTR prediction. The proposed DINMP consists of three part. First, a non-sequential approach is employed to represent the behaviors sequence  in a sparse  key-vector format.  The vector contains rich behavior info such as time, frequency, and category. Moreover, a novel time-aware attention mechanism takes such rich information into account mining deep interest of the user. In addition, a multi-dimensional partition module is introduced, where behaviors are partitioned into different time buckets to capture interaction in different time. We also argued that behaviors can also be partitioned into different categories and interactions among these categories can be captured. At last, experiment results demonstrate the effectiveness of our models both on advertising and recommender datasets.

It is worth mentioning in this work that the sparse key-vector format can represent more valuable data in a compact way. We believe this non-sequential approach opens a new direction for modeling extremely long-term user behavior data. In the future, we will push forward the research on learning algorithms as well as the online serving system.

% \section*{Acknowledgments}
% \appendix
% \section{\LaTeX{} and Word Style Files}\label{stylefiles}

%% The file named.bst is a bibliography style file for BibTeX 0.99c

% \newpage
% \bibliographystyle{named}
\bibliographystyle{ACM-Reference-Format}
\bibliography{kdd}

\end{document}